\documentclass[twocolumn,superscriptaddress,aps,prl]{revtex4}
\usepackage{amsmath}
\usepackage{graphicx}
\usepackage{dcolumn}
\usepackage{color}
\usepackage{ulem}
\usepackage{bm}
\usepackage{braket}
\usepackage[colorlinks,citecolor=blue]{hyperref}

\begin{document}

\title{Composite Spin Approach to the Blockade Effect in Rydberg Atom Arrays}
\author{Lei Pan}
\affiliation{Institute for Advanced Study, Tsinghua University, Beijing,100084, China}
\author{Hui Zhai}
\email{hzhai@tsinghua.edu.cn}
\affiliation{Institute for Advanced Study, Tsinghua University, Beijing,100084, China}
\date{\today}

\begin{abstract}
The Rydberg blockade induces strongly correlated many-body effects in Rydberg atom arrays, including rich ground-state phases and many-body scar states in the excitation spectrum. In this letter, we propose a composite spin representation that can provide a unified description for major features in this system. The composite spin combines Rydberg excitation and the auxiliary fermions, which are introduced to implement the Rydberg blockade constraint automatically. First, we focus on the PXP model describing one-dimensional arrays with the Rydberg blocking radius being a lattice spacing. Using composite spins, the ground state is simply a ferromagnetic product state of composite spins, and the magnon excitations of these composite spins can accurately describe the many-body scar states and signal the quantum phase transition driven by detuning. 
Then, we show that Rydberg atom arrays with different blocking radii can share a universal description of the composite spin representation, and the difference between different blockade radii can be absorbed in the formation of composite spins.  
\end{abstract}

\maketitle

The Rydberg atom arrays have emerged as a new frontier to study strongly correlated quantum matters \cite{review1,review2,review3}. Optical tweezers freeze the spatial motion of atoms, and coherent lasers selectively couple neutral atoms between its ground state and one of the Rydberg states, making the system effectively described by a spin-$1/2$ lattice model. The Rydberg blockade effect introduces strong interactions between spins. In this system, a series of experiments have realized novel quantum phases, and the critical behaviors in phase transitions between them \cite{exp1,exp2,exp3}. These experiments have also discovered intriguing quantum many-body dynamics \cite{exp1,exp4,exp5,exp6,exp7,exp8}, especially the partial violation of quantum thermalization phenomenon known as quantum many-body scars \cite{exp1,PXP_theory,NP_papic}. These experiments have attracted considerable theoretical attention to investigating the many-body physics in this model \cite{PXP_theory,NP_papic,PRB_Papic,PXP_theory2,PXP_theory3,PXP_theory4,PXP_theory5,PXP_theory6,PXP_theory7,PXP_theory8,PXP_theory9,PXP_theory10,related2,PXP_theory11,PXP_theory12,PPXPP1,PPXPP2,yao,Cheng}.  

The strong blockade effect within the blockade radius makes the model a strongly correlated one. A key ingredient to treat strongly correlated models is finding a proper ``quasi-particle" that includes the dominative correlation effect. Then, the system behaves as a weakly interacting one under this quasi-particle representation. One canonical example is the composite fermions by attaching magnetic flux to fermions. The fractional quantum Hall effect, including both the ground state and excitations, then acquires a simple and unified description with the help of the composite fermions \cite{composite-fermion}. Other examples include the slave particles for quantum spin liquids in frustrated spin models \cite{slave_particle} and the dressed impurity cloud for the Kondo effect \cite{Kondo}. 

\begin{figure}[t]
    \centering
    \includegraphics[width=0.45\textwidth]{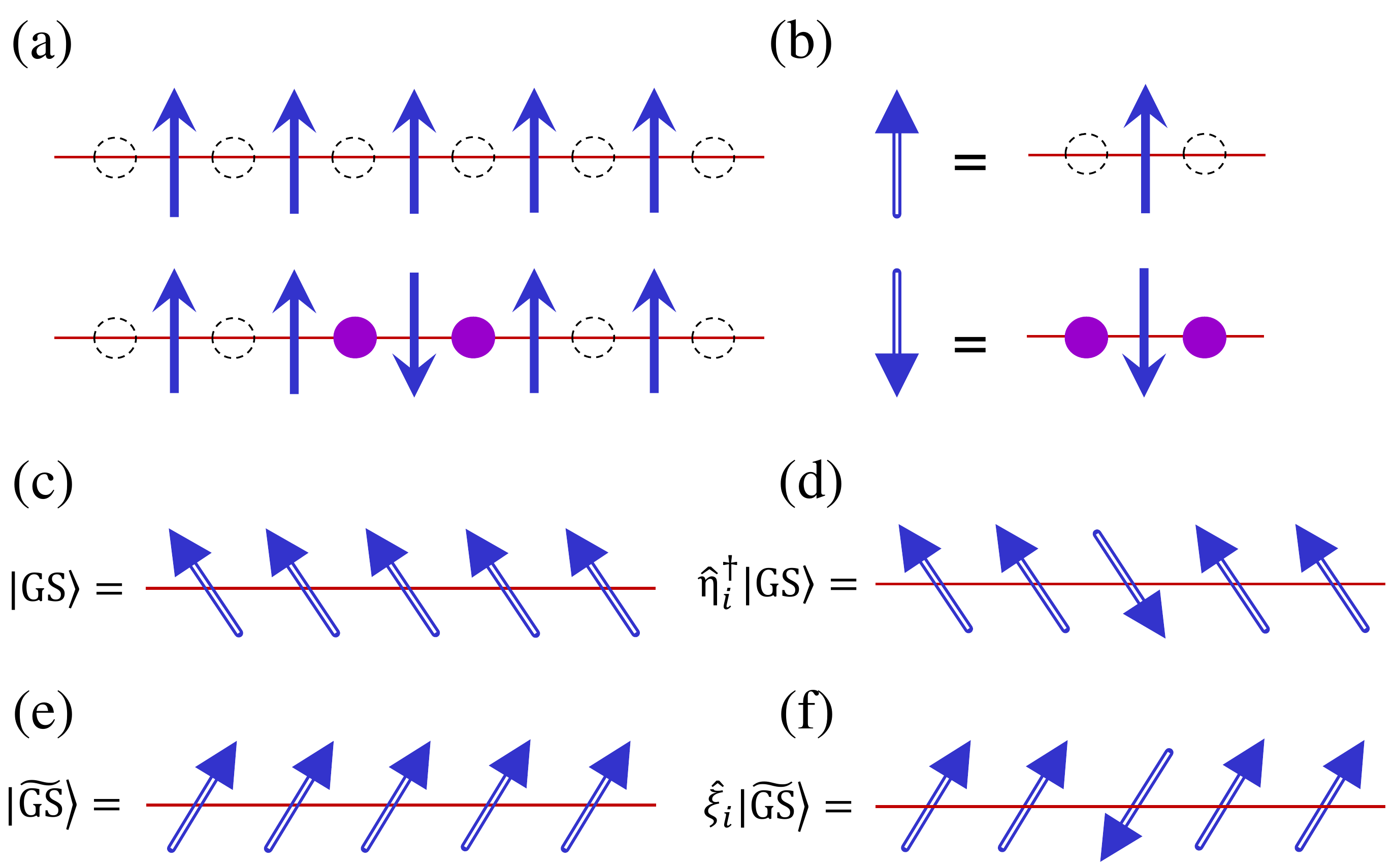}
    \caption{(a) Schematic of the Rydberg blockade PXP model and its auxiliary fermion representation. (b) Schematic of the composite spins. (c) illustrates the ground state as polarized ferromagnetic state of composite spins, and (d) illustrates the magnon excitations of composite spins. When $\Delta=0$, the positive energy states and the negative energy states of the PXP model are related by a unitary transition. (e) is schematic of the highest energy state, and (f) lowers energy from (e) and generates state corresponding to (d).  }
     \label{composite-spin}
\end{figure}

In this letter, we present a composite spin representation for describing the Rydberg blockade-induced many-body effects in the Rydberg atom arrays. We introduce auxiliary fermions, because the Pauli exclusion principle of these fermions can naturally implement the constraints from the Rydberg blockade \cite{Cheng}. We combine the fermions with the original spins to form the composite spins. To demonstrate how it works, we first focus on a one-dimensional array and consider the Rydberg blockade radius being one lattice spacing, which can be described by the so-called PXP model. We show that the composite spin representation provides a simple and unified description of main physics in the PXP model. The main results are following:

1) The ground state is a polarized ferromagnetic state of the composite spins. 

2) The many-body scar states are well described by a set of magnon excitations of these composite spins. 

3) The quantum phase transition driven by detuning is signaled by the magnon excitation being gapless. 

We also show that such a description can be extended to cases in which the Rydberg blockade radius is multiple lattice spacing.

\textit{Composite Spins.} For each Rydberg atom, the laser couples the atom from its ground state (denoted by $\ket{\uparrow}$ here) and a Rydberg state (denoted by $\ket{\downarrow}$), with the coupling strength $\Omega$ and the detuning $\Delta$. And the Hamiltonian is given by 
\begin{equation}
\hat{H}=\sum\limits_{i}\left(\Omega \hat{S}_i^x+\Delta \hat{S}^z_i\right). \label{Hamitonian}
\end{equation}  
In addition, the Rydberg blockade effect imposes a constraint that there cannot be more than one Rydberg atom within the Rydberg radius. When the blockade radius is one lattice spacing, the constraint prevents any two neighboring spins being both $\ket{\downarrow}$, and it requires that for all physical states and for all sites $i$, 
\begin{equation}
\left(\frac{1}{2}-\hat{S}^z_i\right)\left(\frac{1}{2}-\hat{S}^z_{i+1}\right)\ket{\Psi}=0. \label{constraint}
\end{equation}
Eq. \ref{Hamitonian} and Eq. \ref{constraint} together gives rise to the PXP model. 

Here we present an alternative way to implement the Rydberg blockade constraint Eq. \ref{constraint} by introducing spinless fermions sitting at each link \cite{Cheng}, as shown in Fig. \ref{composite-spin}(a). We write the model as  
\begin{equation}
\hat{H}=\sum\limits_{i}\left[\frac{\Omega}{2} \left(\hat{S}_i^{-}f^\dag_{i-1,i}\hat{f}^\dag_{i,i+1}+\hat{S}^{+}_i\hat{f}_{i,i+1}\hat{f}_{i-1,i}\right)+\Delta \hat{S}^z_i\right], \label{Hamitonian_gauge}
\end{equation} 
where $\hat{f}^\dag_{i,i+1}$ ($\hat{f}_{i,i+1}$) is the fermion creation (annihilation) operator at the link between site $i$ and $i+1$. This model possesses local $U(1)$ gauge symmetry $\hat{f}_{i,i+1}\rightarrow e^{i\theta_i}\hat{f}_{i,i+1}$, $\hat{S}^{+}_i\rightarrow e^{-i\theta_i}\hat{S}^{+}_i$ and $\hat{S}^{+}_{i+1}\rightarrow e^{-i\theta_i}\hat{S}^{+}_{i+1}$, and it gives rise to local conserved gauge charge $Q_{i,i+1}=S^z_{i}+S^z_{i+1}+n_{i,i+1}$, where $n_{i,i+1}$ is the fermion number. When we focus on the gauge sector with all $Q_{i,i+1}=1$, it is clear that when the operator $\hat{S}^{-}_i$ flips $\ket{\uparrow}$ state at site $i$ to $\ket{\downarrow}$ state, it simultaneously creates two fermions at two neighboring links between site $i$ and site $i-1$ (and site $i+1$). Hence, spin flip can no longer occur in site $i-1$ and site $i+1$ anymore. Hence, we can introduce the composite spins by combining a pair of fermions with each spin. That is to say, for each site, we define composite spin $\tau_i$ as 
\begin{align}
&\ket{\Uparrow}=\ket{\uparrow}\otimes|00\rangle,\\
&\ket{\Downarrow}=\ket{\downarrow}\otimes|11\rangle,
\end{align}
and we schematically illustrate the composite spins in Fig. \ref{composite-spin}(b).

\begin{figure}[t]
    \centering
    \includegraphics[width=0.45\textwidth]{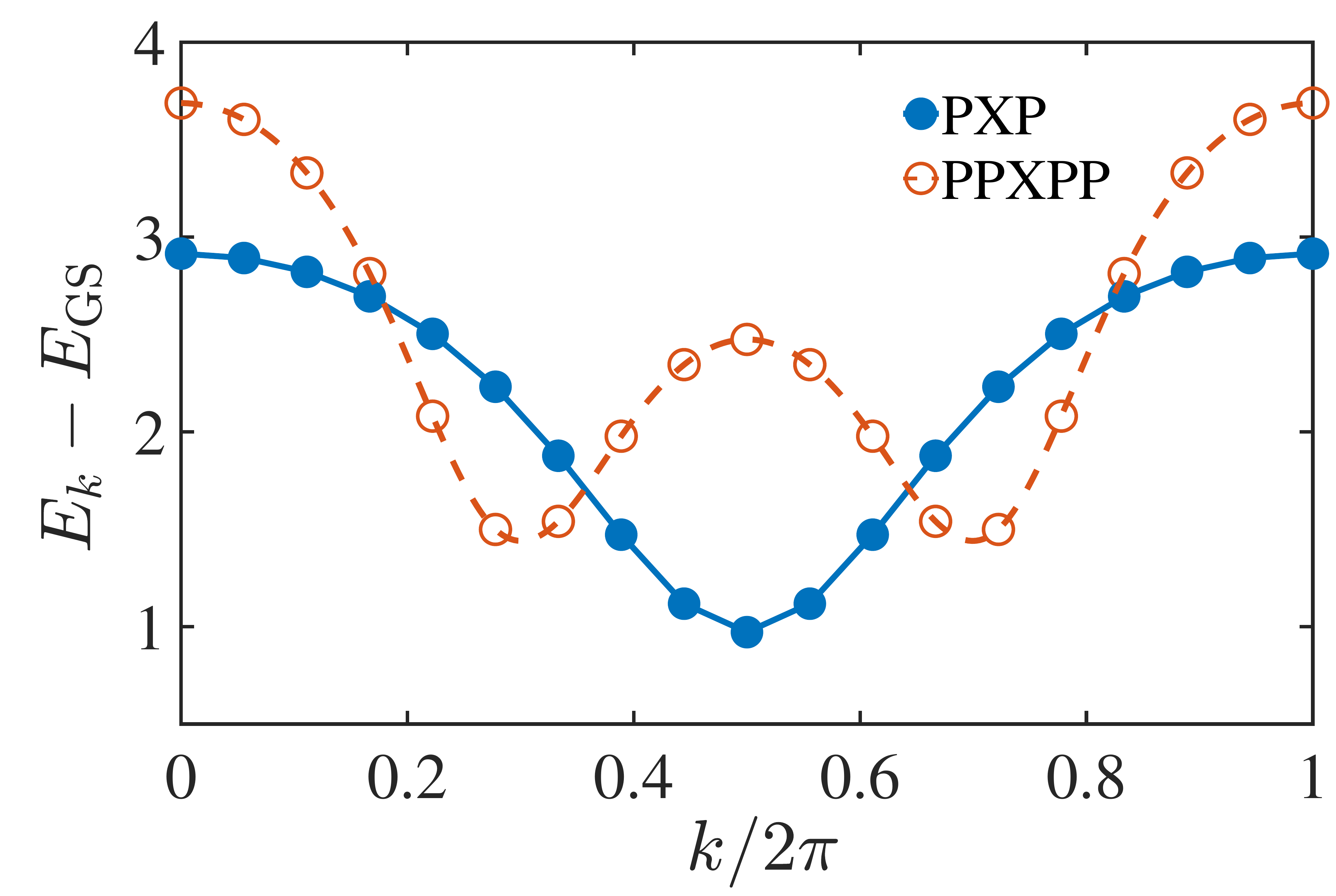}
    \caption{The excitation energy of the single magnon excitation as a function of momentum. The solid blue dots with a solid line is for the PXP model (blockade radius being one lattice spacing), and the red open dots with a dashed line is for the PPXPP model (blockade radius being two lattice spacing). Here the system size of the plot is for $18$ sites.}
     \label{spin-wave}
\end{figure}

\textit{Ground State.} In term of the composite $\tau$-spin, the Hamiltonian simply reads 
\begin{equation}
\hat{H}=\sum\limits_{i}\left(\Omega \hat{\tau}_i^x+\Delta \hat{\tau}^z_i\right). \label{Hamitonian-tau}
\end{equation} 
Note that we do not need to add the Rydberg blockade constraint when the model is written in terms of $\tau$-spin, because the Fermi exclusion principle automatically forbids the composite spins in two neighboring sites being both $\ket{\Downarrow}$, implementing the constraint from Rydberg blockade effect. Motivated by Eq. \ref{Hamitonian-tau}, it is natural to speculate that the ground state is a product state of composite spins polarized by an external field, i.e. 
\begin{equation}
\ket{\text{GS}}=\frac{1}{\mathcal{N}}\prod_{i}\big(u\ket{\Uparrow}_i+v\ket{\Downarrow}_i\big), \label{GS}
\end{equation}
where $\mathcal{N}$ is the normalization factor. This state is illustrated by Fig. \ref{composite-spin}(c). Below we will set $u=1$, and then $v$ is the unique parameter in the wave function. 

We first consider the situation $\Delta=0$. The parameter $v$ is determined by minimizing  $E_\text{GS}=\bra{\text{GS}}\hat{H}\ket{\text{GS}}$ and we obtain $v=-0.636$. We also compare the wave function with the ground state obtained from exact diagonalization (ED), where the overlap between them is as high as $0.99$. This shows that the wave function Eq. \ref{GS} provides a very precise description of the ground state. Moreover, note that when $\Delta=0$, we can define a unitary transformation $\hat{U}=\prod_{i}\sigma^z_i$, and $\hat{U}\hat{H}\hat{U}=-\hat{H}$. Thus, if $\ket{\Psi}$ is an eigenstate of $\hat{H}$ with negative eigen-energy $E$, $\hat{U}\ket{\Psi}$ generates an eigenstate with positive eigen-energy $-E$. Thus, $\ket{\widetilde{\text{GS}}}=\hat{U}\ket{\text{GS}}$ generates the highest energy state, which takes the same form as Eq. \ref{GS} but $v=0.636$. This state is illustrated by Fig. \ref{composite-spin}(e).

\begin{figure}[t]
    \centering
    \includegraphics[width=0.38\textwidth]{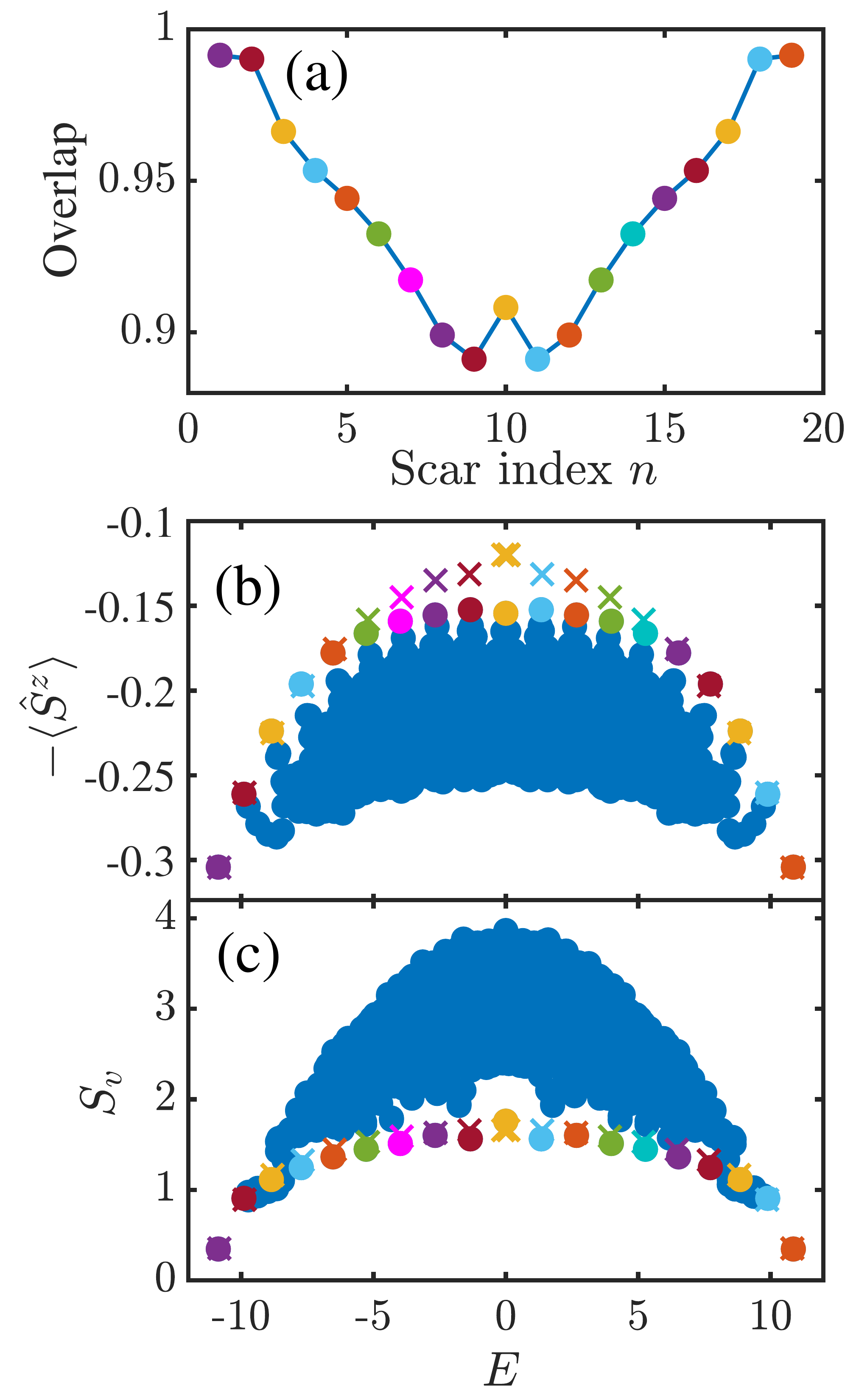}
    \caption{(a) Overlap between the many-body scar states and the quantum states constructed from the $\pi$-magnon excitations. These states are marked in (b) and (c) with the same colors. (b-c) The spin polarization $-\langle\hat{S}_z\rangle$ along $\hat{z}$ and the entanglement entropy $\mathcal{S}_\text{v}$ between the left and the right half systems. The dots represent results from ED with system size $N_\text{s}=18$. The crosses with corresponding colors represent results predicted by $\pi$-magnon states. Here we consider $\Delta=0$ and energy is in unit of $\Omega$.     }
     \label{scar}
\end{figure}

\textit{Magnon Excitation.} Since we have obtained a very accurate wave function for the ground state, now we consider excitations above it. The most natural excitation is the magnon or spin wave. We define the operator $\hat{\eta}_i$ that only acts on $\tau$-spin at site $i$ as
\begin{equation}
\hat{\eta}^\dag_i(u\ket{\Uparrow}_i+v\ket{\Downarrow}_i)=v\ket{\Uparrow}_i-u\ket{\Downarrow}_i, \label{eta}
\end{equation} 
which is illustrated in Fig. \ref{composite-spin}(d) and satisfies the condition $(\hat{\eta}^\dag_i)^2=0$. Then, we consider the magnon excitation with momentum $k$, and up to a normalization factor, it is defined as 
\begin{equation}
\ket{k}\propto\hat{\eta}^\dag_{k}\ket{\text{GS}}=\frac{1}{N_\text{s}}\sum\limits_{i}e^{ik R_i}\hat{\eta}^\dag_i\ket{\text{GS}}, \label{magnon}
\end{equation}
where $N_\text{s}$ is the number of total sites. We compute the energy $E_k=\bra{k}\hat{H}\ket{k}$, and plot $E_k-E_\text{GS}$ as the single magnon dispersion in Fig. \ref{spin-wave}. We can see that for the PXP model, the single magnon excitation displays a minimum at momentum $\pi$ and acquires a rather large gap.  

\begin{figure}[t]
    \centering
    \includegraphics[width=0.50\textwidth]{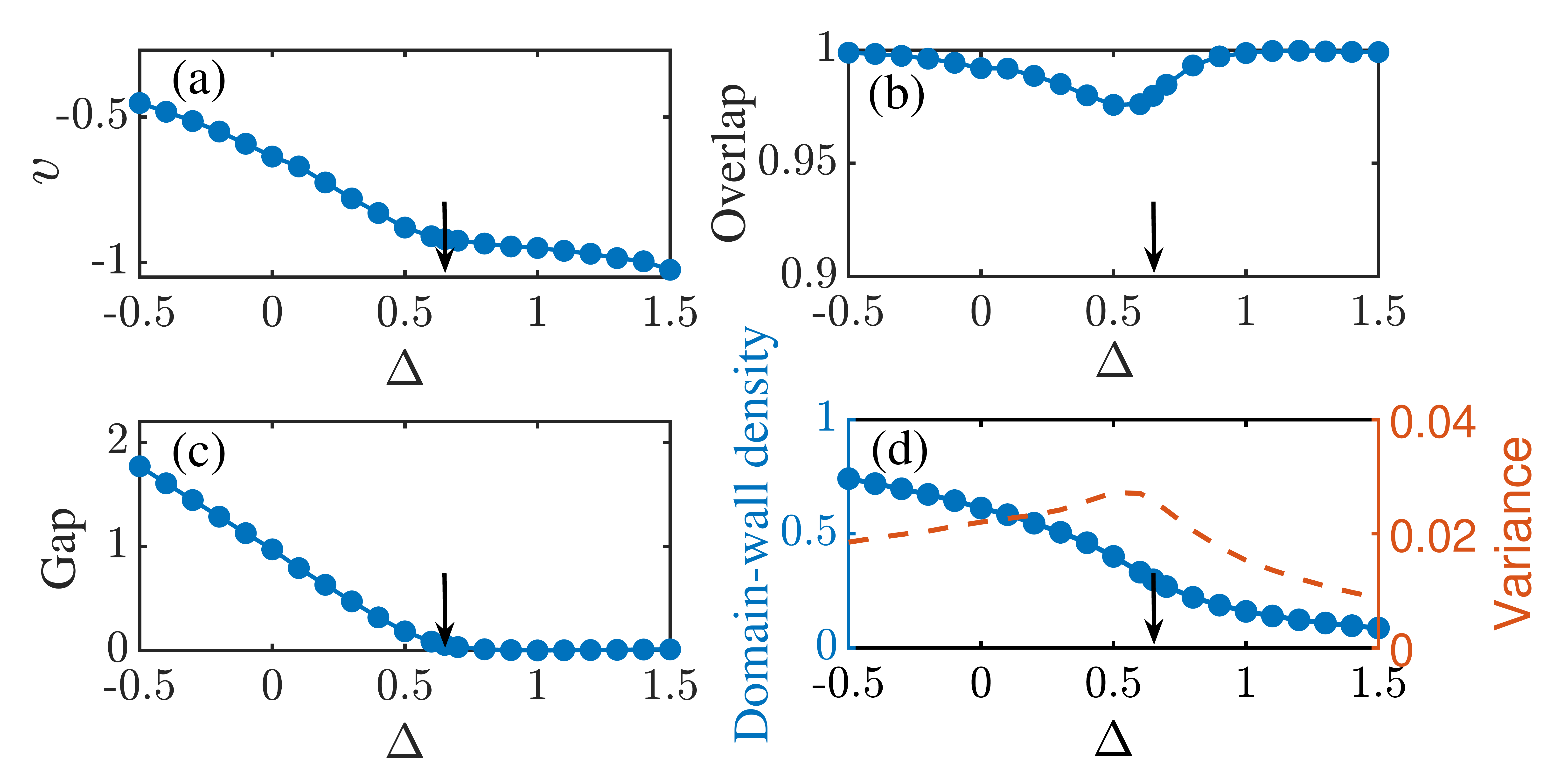}
    \caption{(a): The optimal $v$ for our magnon wave function. (b): The overlap between our magnon wave function and the ground state obtained by ED with system size $N_\text{s}=18$. (c) The gap predicted by our magnon wave function. (d) The domain wall density and variance for the ground state predicted by our magnon wave function. Arrows label the locations of expected quantum phase transition. $\Delta$ is in unit of $\Omega$. Here all data shown are for $N_{\text s}=18$.}
     \label{phase_transition}
\end{figure}

We can then define another operator $\hat{\xi}_i$ that only acts on $\tau$-spin at site $i$ as 
\begin{equation}
\hat{\xi}_i(u\ket{\Uparrow}_i-v\ket{\Downarrow}_i)=v\ket{\Uparrow}_i+u\ket{\Downarrow}_i,
\end{equation} 
which is illustrated in Fig. \ref{composite-spin}(f) and satisfies the condition $\hat{\xi}_i^2=0$. We can also introduce 
\begin{equation}
\ket{\tilde{k}}\propto\hat{\xi}_{k}\ket{\widetilde{\text{GS}}}=\frac{1}{N_\text{s}}\sum\limits_{i}e^{ik R_i}\hat{\xi}_i\ket{\widetilde{\text{GS}}},
\end{equation}
and it is easy to see that $\ket{\tilde{k}}=\hat{U}\ket{k}$ with the unitary transformation $\hat{U}$ defined above.

\textit{Many-body Scar States.} Because of the composite nature, the $\tau$ spins are not independent because two neighboring $\tau$ spins do not commute. The residual interactions between magnons can lead to thermalization of these excitations. For instance, Beliaev damping can occur for magnons away from momentum-$\pi$. Therefore, we focus on the magnon with momentum-$\pi$. Because this magnon sits at the bottom of excitation spectrum and possesses a gap from the ground state, it is more immune to quasi-particle interactions and is more likely to violate thermalization.   

Motivated by this insight, we construct a set of orthogonal states composed of 
\begin{equation}
\{\ket{\text{GS}}, \hat{\eta}^\dag_{\pi}\ket{\text{GS}}, (\hat{\eta}^\dag_{\pi})^2\ket{\text{GS}},\dots, (\hat{\eta}^\dag_{\pi})^{N_\text{s}/2}\ket{\text{GS}}  \}, \label{pi-magnon}
\end{equation}
where $\ket{\text{GS}}$ is the ground state Eq. \ref{GS} with $u=1$ and $v=-0.636$ determined by energy minimization. This set of states includes a different number of magnons from zero to $N_\text{s}/2$, resembling the nearly equal energy spacing between different many-body scar states in the PXP model. Therefore, we can establish a one-to-one correspondence between this set of states and the negative energy scar states. For positive energy scar states, we construct another set of orthogonal states composed of 
\begin{equation}
\left\{\ket{\widetilde{\text{GS}}}, \hat{\xi}_{\pi}\ket{\widetilde{\text{GS}}}, \hat{\xi}_{\pi}^2\ket{\widetilde{\text{GS}}},\dots, \hat{\xi}_{\pi}^{N_\text{s}/2}\ket{\widetilde{\text{GS}}}  \right\},
\end{equation}
which are respectively related to states in Eq. \ref{pi-magnon} by $\hat{U}$ discussed above.

Remarkably, as shown in Fig. \ref{scar}, we find high overlaps for all these states \textit{without using any fitting parameter}. In Fig. \ref{scar}(b) and (c), we plot the average of spin polarization $\langle \hat{S}^z\rangle$ and the bipartite entanglement entropy $\mathcal{S}_\text{v}$, where $\hat{S}^z=\frac{1}{N_{s}}\sum_{i}\hat{S}^z_i$ is the average of spin along $\hat{z}$, and the bipartite entanglement entropy is obtained by dividing the system equally into the left and right halves. We compare our prediction of the $\pi$-magnon wave functions with the ED results. Fairly good agreements are reached, especially for these states near the bottom and the top of the energy spectrum. The overlaps slightly decrease, and the degrees of agreement for $\langle \hat{S}^z\rangle$ are slightly reduced when energy increases (or decreases) toward the middle of the spectrum. This is because the total magnon number increases when energy increases, and the effect of residual interactions between magnons also becomes pronounced.

\textit{Phase Transition.} It is known that the PXP model displays a quantum phase transition to a $Z_2$ symmetry breaking phase as $\Delta$ increases \cite{Sachdev,Fendley,Zoller}. When $\Delta=0$, the magnon excitation acquires a sizable gap, and mixing between states with different magnon numbers is negligible. However, as $\Delta$ increases toward the phase transition, the magnon gap is significantly reduced and approaches zero, and the coupling between states within the set Eq. \ref{pi-magnon} can no longer be ignored. Thus, we will slightly modify our scheme. We still focus on the subspace spanned by Eq. \ref{pi-magnon}. However, instead of fixing $v=-0.636$ to minimize $\bra{\text{GS}}\hat{H}\ket{\text{GS}}$, we first keep $v$ as an undetermined parameter and diagonalize the Hamiltonian in this subspace. The diagonalization procedure mixes states with different magnon numbers, and generates a new ground state $\ket{\text{GS}^\star}$ with energy $E_{\text{GS}^\star}(v)$. The optimum $v$ is determined by minimizing $E_{\text{GS}^\star}(v)$ for any given $\Delta$.

\begin{figure}[t]
    \centering
    \includegraphics[width=0.40\textwidth]{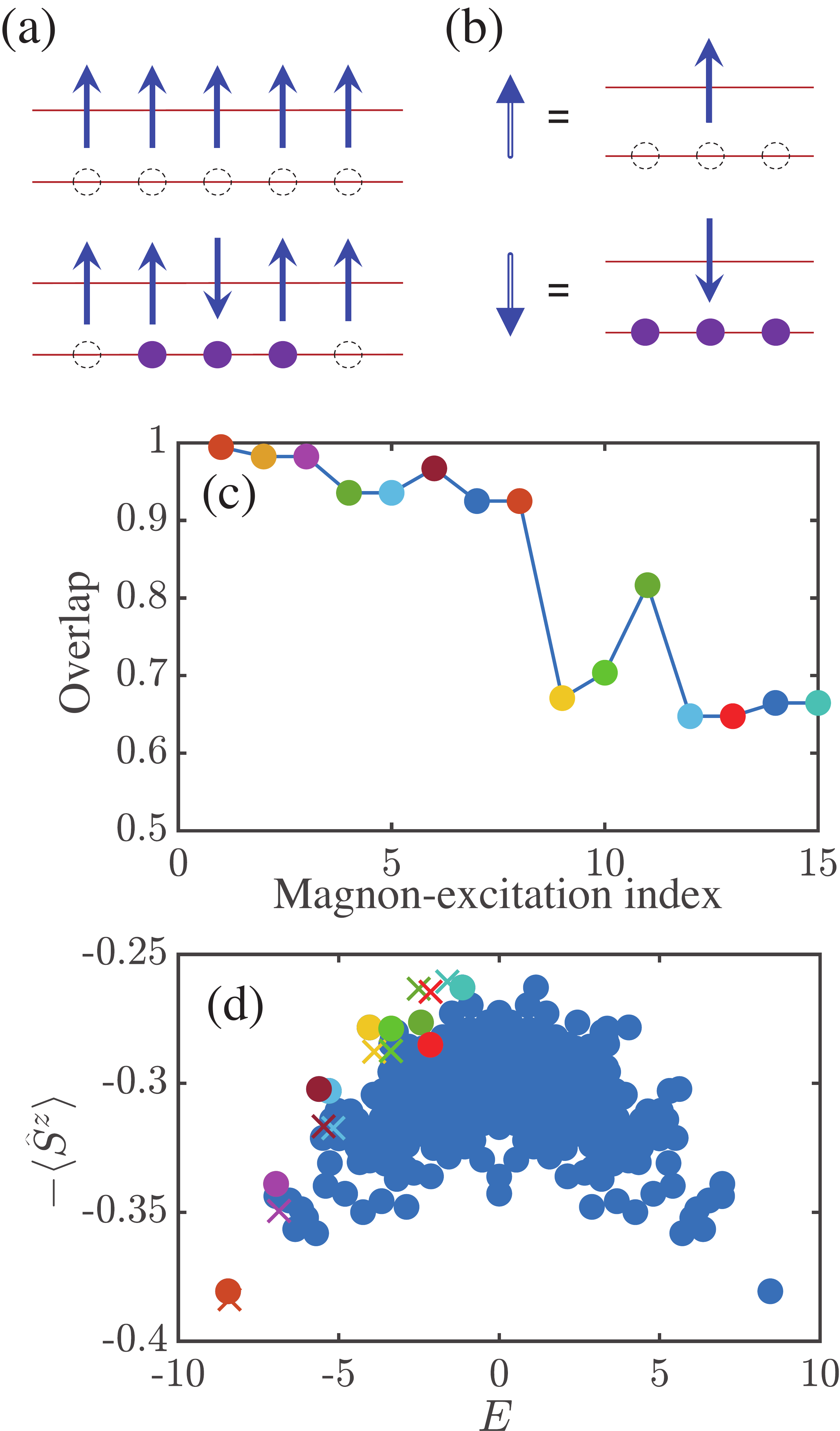}
    \caption{(a-b) Schematic of the auxiliary fermion representation and the composite spins for the PPXPP model. (c-d) The overlap (c) and comparison of $-\langle\hat{S}_z\rangle$ (d) between a set of magnon excitation states and eigenstates obtained by ED. In (d), dots are results from ED, and crosses are obtained by magnon wave function. The same color is used for same states between (c) and (d). Here we keep the total number of magnons up to $4$. All
 plots are for $N_{\text s}=18$. }
     \label{PPXPP}
\end{figure}

Fig. \ref{phase_transition}(a) shows the optimal $v$ determined in this way from which we can see that $v$ is always negative, and $|v|$ keeps increasing as $\Delta$ increases. It displays a kink around the expected critical point $\Delta_\text{c}=0.655$. With the optimal $v$, $\ket{\text{GS}^\star}$ always acquires a large overlap ($>0.97$) with the ground state obtained by ED for all range of $\Delta$, as shown in Fig. \ref{phase_transition}(b). Most remarkably, we find that the excitation gap within this subspace becomes vanishing small when $\Delta>\Delta_\text{c}$, signaling the symmetry-breaking nature of the quantum phase transition. Note that both many-body scars states and the quantum phase transition in the PXP model can be well captured by the same set of $\pi$-magnon excitations, which is consistent with the previous study revealing the connection between scar states at $\Delta=0$ and the low-energy critical states around $\Delta_\text{c}$ \cite{yao}.

Ref. \cite{exp1} measured the so-called ``domain-wall" density and its variance. When two neighboring atoms are both in the ground states, it is counted as a domain-wall of the $Z_2$ symmetry breaking state. We compute these two quantities using the wave function $\ket{\text{GS}^\star}$, as shown in Fig. \ref{phase_transition}(d). We find that the domain-wall density monotonically decreases as $\Delta$ increases, and its variance displays a peak around $\Delta_\text{c}$. This feature also qualitatively agrees with the experimental observation \cite{exp1}. 

\textit{PPXPP Model.} The composite spin picture can be easily generalized to atom arrays with different Rydberg blockade radii and geometries. For instance, we consider a one-dimensional array with the Rydberg blockade radius being two lattice spacing. That is to say, not only two nearest spins but also two next nearest spins cannot be both in $\ket{\downarrow}$ state. In addition to the constraint Eq. \ref{constraint}, we need to apply another constraint
\begin{equation}
\left(\frac{1}{2}-\hat{S}^z_i\right)\left(\frac{1}{2}-\hat{S}^z_{i+2}\right)\ket{\Psi}=0. \label{constraint2}
\end{equation}
This leads to a so-called PPXPP model \cite{PPXPP1,PPXPP2}. These constraints can also be taken care of automatically by the Pauli exclusion principle by introducing auxiliary fermions sitting at sites (instead of links in the case of the PXP model), as shown in Fig. \ref{PPXPP}(a). And we rewrite the model as  
\begin{equation}
\hat{H}=\sum\limits_{i}\left[\frac{\Omega}{2} \left(\hat{S}_i^{-}f^\dag_{i-1}\hat{f}^\dag_{i}\hat{f}^\dag_{i+1}+\hat{S}^{+}_i\hat{f}_{i+1}\hat{f}_{i}\hat{f}_{i-1}\right)+\Delta \hat{S}^z_i\right]. \label{Hamitonian_gauge-2}
\end{equation} 
It is easy to see that the spin flip at site $i$ simultaneously creates three fermions, which forbids further spin flips at the nearest and the next nearest sites. Hence, as shown in Fig. \ref{PPXPP}(b), we modify the definition of the composite spin $\tau_i$ into 
\begin{align}
&\ket{\Uparrow}=\ket{\uparrow}\otimes|000\rangle,\\
&\ket{\Downarrow}=\ket{\downarrow}\otimes|111\rangle.
\end{align} 
Such an auxiliary fermions and composite spin description can also be generalized to a one-dimensional array with even larger blockade radius.  

Under the composite spin representation, the Hamiltonian still reads as Eq. \ref{Hamitonian-tau}, and the ground state ansatz still reads as Eq. \ref{GS}. When $\Delta=0$, we obtain $v=-0.498$ by minimizing $\bra{\text{GS}}\hat{H}\ket{\text{GS}}$, and the overlap between this state and the ground state obtained by ED is also as high as $0.99$. The magnon excitation is also generated by Eq. \ref{eta} and Eq. \ref{magnon}. The single magnon dispersion is also shown in Fig. \ref{spin-wave}, which shows the minima of single magnon dispersion located nearby momentum $\approx 2\pi/3$ and $\approx 4\pi/3$. Following the same stratagem discussed in the PXP model, we consider a subspace spanned by 
\begin{equation}
\{\ket{\text{GS}}, \hat{\eta}^\dag_{\frac{2\pi}{3}}\ket{\text{GS}}, \hat{\eta}^\dag_{\frac{4\pi}{3}}\ket{\text{GS}}, \hat{\eta}^\dag_{\frac{2\pi}{3}}\hat{\eta}^\dag_{\frac{4\pi}{3}}\ket{\text{GS}},\dots \},
\end{equation} 
and we construct a set of orthogonal states in this subspace and compare them with the eigenstates obtained by ED. The wave function overlaps are shown in Fig. \ref{PPXPP}(c), and the corresponding eigenstates are marked by the same colors in Fig. \ref{PPXPP}(d). We also compare the prediction of spin polarization between these magnon states and the results from ED. These states also show the most significant deviation from the thermal value in a finite-size system. Like the PXP model, the agreements also decrease as energy increases toward the middle of the spectrum because the effect of the quasi-particle interaction becomes stronger as the magnon number increases. Overall, the agreements are not as perfect as the PXP model because the PPXPP model confronts more constraints that induce stronger interaction between quasi-particle.   

\textit{Summary.} The analytical wave function ansatz presented by this work can provide a unified description of both the ground state and the non-thermal excited states in the PXP model and can also provide a universal description of models with different blockade radii. We acknowledge that certain aspects of our wave function have also been discussed in previous works. The same ground state has been studied in Ref. \cite{related,related2,PRB_Papic}, and a similar but different $\pi$-magnon wave function has been proposed for many-body scar states \cite{PXP_theory6}. However, our work provides a so-far the most systematical and unified view for Rydberg blockade-induced many-body phenomena, and the validity of our approach is supported by the high overlaps between our wave function ansatz and the wave functions obtained by ED. This wave function ansatz can be extended to study high-dimension models and the non-equilibrium dynamical processes. The correlation effects imposed by the Rydberg blockade are automatically taken care of by combining fermions with spins. Therefore, the composite spins serve as a natural building block, resulting in an intuitive physical picture. 

\textit{Acknowledgment.}
The project is supported by Beijing Outstanding Young Scholar Program, NSFC Grant No.~11734010 and the XPLORER Prize.

\end{document}